\documentclass[12pt,a4paper]{article}
\usepackage{axodraw,cite}
\makeatletter
\def\reportno#1{\gdef\@reportno{#1}\let\ps@oldplain\ps@plain
  \gdef\ps@plain{\global\let\ps@plain\ps@oldplain\let\@mkboth\@gobbletwo
    \def\@oddhead{\parbox{\textwidth}{\raggedleft\@reportno}}}}
\let\@reportno\@empty
\makeatother
\evensidemargin\oddsidemargin
\advance\topmargin  by-0.5in
\advance\textheight by 0.5in
\def\bra#1{\left\langle#1\right|\;\!}           % bra: <#|
\def\ket#1{\;\!\left|#1\right\rangle}           % ket: |#>
\def\slashed#1{\rlap{/}#1}                      % slash (/)
\def\Slashed#1{\rlap{\kern0.06em/}#1}           % slash on upper case
\def\eg{\emph{e.g.}}
\def\etal{\emph{et al}}                         % period to add by hand
\def\etc{\emph{etc.}}
\def\ie{\emph{i.e.}}
\title{Factorisation in Higher-Twist
  \\
  Single-Spin Amplitudes
  \\[3pt]
  \small (submitted to the \emph{European Physical Journal\/}
  \textbf{C})}

\reportno{EPTCO-98-001}

\author{Philip G. Ratcliffe\,\thanks{E-mail: pgr@fis.unico.it}
  \\
  \small\emph{Dipartimento di Scienze, Universit\`a
    dell'Insubria---sede di Como,}
  \\
  \small\emph{via Lucini 3, 22100 Como, Italia}
  \\
  \small and
  \\
  \small\emph{Istituto Nazionale di Fisica Nucleare---sezione di
    Milano, Italia}}

\date{revised November 1998}
\begin{document}
\maketitle
%---------------------------------- Abstract ----------------------------------%
\begin{abstract}

We analyse the twist-three amplitudes that can give rise to single-spin
asymmetries in hadron-hadron scattering; in so doing we bring to light a
novel factorisation property. As already known, the requirement of an
imaginary part leads to consideration of twist-three contributions that
are also related to transverse spin in deep-inelastic scattering. In
particular, when an external line becomes soft in contributions arising
from three-parton correlators, the imaginary part of an internal
propagator may be exposed. As shown here, it is precisely this
kinematical configuration that permits the factorisation. An important
feature is the resulting simplification: the calculation of tens of
Feynman diagrams normally contributing to such processes is reduced to
the evaluation of products of the simple factors derived here and known
two-body helicity amplitudes. We thus find clarifying relations between
the spin-dependent and spin-averaged cross-sections and formulate a
series of selection rules. In addition, the kinematical dependence of
such asymmetries, is rendered more transparent.


\end{abstract}
\newpage
%--------------------------------- Main Body ----------------------------------%
\section{Introduction: single-spin phenomenology}

A large body of information has now been gathered in regard of
single-spin asymmetries in semi-inclusive hadronic processes
\cite{DataG2}, where the striking feature is the magnitude of such
effects (up to $\sim$40\%). Such phenomena present a theoretical
challenge: to find sizeable interfering spin-flip and non-flip
amplitudes with relative imaginary phases, a severe difficulty for a
gauge theory with near-massless fermions \cite{KPR}. At the same time,
although subject to some early confusion, there is now a clear
understanding of the nature and r\^ole of three-parton twist-three
correlators in the transverse-spin dependence of deep-inelastic
scattering (DIS) \cite{Twist3,ET-1,PGR-1}. However, the distribution
functions associated with such structures will be difficult to study
comprehensively \cite{RLJ}, especially if consideration is restricted to
DIS. Indeed, although data are steadily becoming available
\cite{DataSSA}, further experimental knowledge will be necessary for a
complete description of transverse-spin phenomena. Thus, single-spin
asymmetries, which are intimately related to the same twist-three
amplitudes, may be an invaluable integration of our knowledge in this
area.

The experimental aspects of single-spin asymmetries are well documented
\cite{KH}: the main point to stress is that the measured effects do not
appear at all suppressed, even for values of $p_T$ where it might be
hoped that perturbative QCD (pQCD) should be applicable. On the other
hand, it has long been held that they would not be reproducible in pQCD
\cite{KPR}, although a satisfactory (but largely incomplete) description
of such asymmetries is provided by a number of non-perturbative
approaches.

One might question whether or not it even makes sense to apply pQCD to
processes that, for the time being, have only been measured at
relatively small values of $p_T$. However, recall that Anselmino \etal.\
\cite{ABM} have made successful fits to the existing pion data, based on
pQCD-inspired models. Moreover, the hyperon data does reach large values
of $p_T$, where there is no hint of the polarisation disappearing. If
these transverse-spin effects do have a common origin, then one might
hope that a perturbative approach should give a \emph{reasonable\/}
description down to some typical hadronic scale. In this respect,
although Teryaev \cite{OVT} has recently shown that twist-four effects
must become important at large parton $x$, where twist-three
contributions would otherwise induce positivity violation owing to their
lower-power dependence on $(1{-}x)$, this is not an argument against the
applicability of pQCD. Rather, it underlines the well-known fact that
while higher twist is important for $x\to1$, there is an intermediate
region where it is negligible even at very low scales. Indeed, just this
type of process, being so-to-speak only \emph{slightly\/} higher twist,
may well provide clues to the transition between regions.

The basic hurdle lies then in the need for spin-flip amplitudes with
relative imaginary phases; in a suitable helicity basis it can be shown
that single transverse-spin effects are related to the imaginary part of
the interference between spin-flip and non-flip amplitudes. Normally, in
a gauge theory, spin-flip can only be generated via fermion masses, and
phases by loop corrections. However, some time ago Efremov and Teryaev
noted \cite{ET-1} that the loop implicit in diagrams containing an extra
partonic leg (arising in higher-twist transverse-spin effects) naturally
leads to an \emph{unsuppressed\/} imaginary part with spin flip. To
understand this, it is helpful to appreciate that the extra loop
(na{\"\i}vely implying higher order in $\alpha_s$) is accompanied by a
large logarithm. Thus, the associated distribution function is to be
considered at the level of the usual leading-order densities. In other
words, at leading-logarithmic level, the usual infinite sum of terms in
$(\alpha_s\,\log{Q^2})^n$ is present; however, \emph{just the very first
term is missing\/} \cite{PGR-2}. In practice, the extra power of
$\alpha_s$ inherent to these contributions is effectively absorbed into
the hadron-parton correlator.

We note in passing that twist is best considered in terms of the power
of $Q^2$ with which a given contribution appears in a hadronic
cross-section \cite{RLJ}: in the single-spin case one expects
asymmetries to behave as
\begin{equation} \label{eq:asym}
  {\cal A} \; \propto \; \frac{\mu p_T}{\mu^2+p_T^2},
\end{equation}
where $\mu$ is some typical hadronic mass scale. Again, Teryaev
\cite{OVT} has discussed how the necessary inclusion of twist four leads
to the form of the denominator in eq.~(\ref{eq:asym}). Thus, the usual
suppression should be observed asymptotically while a roughly linear
dependence is expected for low values of $p_T$. The intriguing
implication of Teryaev's analysis is that the point of maximum asymmetry
should indicate the onset of the regime dominated by leading twist. If
the hyperon data is typical then this already occurs at around 1\,GeV
for intermediate values of $x$. However, the $p_T$ dependence would
suggest that at the point where higher twist is reduced by a factor 10
the asymmetry will still be ${\sim}30\%$ of its maximum value.

Much progress has been made in the direction of interrelating the
various aspects of polarisation phenomenology
\cite{ET-1,QS-1,KT,HES,Drell-Yan,Pion,XJ}. In particular, in the case of
twist-three contributions, the possibility that one of the
hard-scattering propagators may generate an imaginary part in the soft
limit has already been exploited as a possible mechanism for the large
asymmetries mentioned above. Early work concentrated on prompt-photon
production \cite{ET-1,QS-1,KT,HES}; other processes that have been
considered are pion production \cite{Pion} and Drell-Yan
\cite{Drell-Yan}.

Here we present a systematic analysis to demonstrate how the requirement
of an imaginary part (and thus a soft internal propagator) greatly
simplifies calculations owing to a novel factorisation property of the
Feynman amplitudes involved. After some preliminary definitions in the
next section and clarification of the spin-flip requirement at the
partonic level, section \ref{sec:factor} contains the main derivation
and results, illustrating how the factorisation arises and the simple
selection rules that follow therefrom. In the concluding section we
present the resulting formal expression for the spin-dependent partonic
cross-sections, together with some discussion.

While the technique presented opens the way to simpler and more rapid
calculation, we do not consider it useful to present yet another
evaluation of any particular process for two reasons: firstly, a model
input for the unknown parton correlators would, in any case, be required
and we have nothing new to add there; and, secondly, many calculations
have already been published (as cited above) and this technique should
not, of course, produce different results.

%------------------------------------------------------------------------------%
\section{Preliminaries and definitions}
\label{sec:prelim}

\begin{figure}[htb]
\begin{center}
%==============================================================================%
% T3DIAGS.AXO ((La)TeX input)                                                  %
% Last modified on 25-FEB-1998                     author: Philip G. Ratcliffe %
%==============================================================================%
\begin{picture}(400,150)(0,0)
\multiput(0,0)(150.5,0){3}{
\Line(  0,141)( 20,141) \Line(100,141)( 80,141)
\Line(  0,139)( 20,139) \Line(100,139)( 80,139)
\Line( 11,140)(  8,143) \Line( 11,140)(  8,137)
\Line( 91,140)( 88,143) \Line( 91,140)( 88,137)
\Oval( 50,140)( 10,30)(0)
\Line( 30,133)( 30,105) \Line( 70,133)( 70,105)
\Line( 30,105)( 70,105)
\Gluon( 70,105)( 70, 65){2}{6}
\Line( 30, 65)( 70, 65)
\Line( 30, 37)( 30, 65) \Line( 70, 37)( 70, 65)
\Oval( 50, 30)( 10,30)(0)
\Line(  0, 31)( 20, 31) \Line(100, 31)( 80, 31)
\Line(  0, 29)( 20, 29) \Line(100, 29)( 80, 29)
\Line( 11, 30)(  8, 33) \Line( 11, 30)(  8, 27)
\Line( 91, 30)( 88, 33) \Line( 91, 30)( 88, 27)
}
\Gluon( 50,130)( 50,105){2}{4}
\Gluon( 30,105)( 30, 65){-2}{6}
\Text( 50, 0)[]{(a)}
\Gluon(200,130)(200,98){2}{4}
\GlueArc(180,105)(20,270,360){2}{4}
\Gluon(180,105)(180, 85){-2}{3}
\Gluon(180, 85)(180, 65){-2}{3}
\Vertex(180,85){1}
\Text(200, 0)[]{(b)}
\Gluon(350,130)(350,63){2}{10}
\GlueArc(330, 70)(20,270,360){2}{4}
\Gluon(330,105)(330, 65){-2}{6}
\Text(350, 0)[]{(c)}
\end{picture}
\caption{\label{fig:t3diags}
  Example contributions to twist-three transverse single-spin
  effects.}
\end{center}
\end{figure}
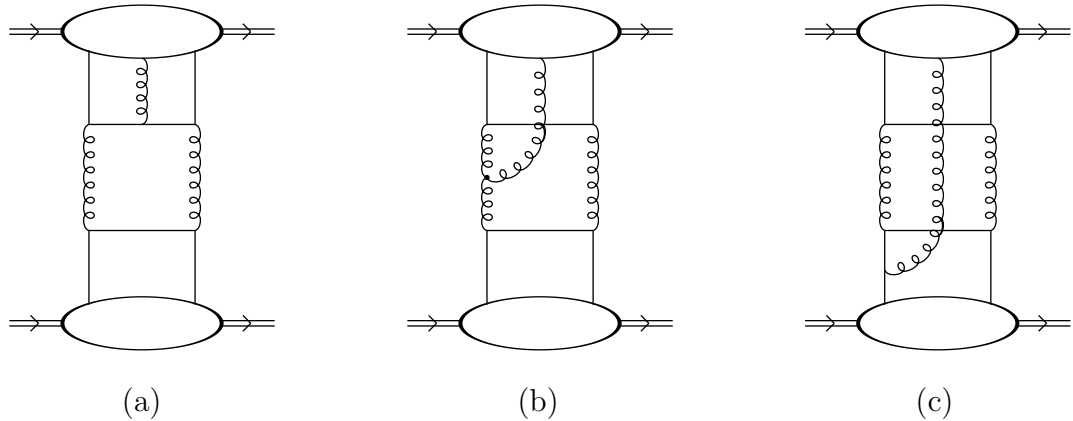

Some relevant twist-three diagrams are displayed in
Fig.\,\ref{fig:t3diags};
such diagrams may contribute to single-spin asymmetries owing to the
imaginary parts implicit in the internal lines, according to the
standard propagator prescription:
\begin{equation}
  \frac{1}{k^2\pm{i}\varepsilon}
  \; = \;
  I\!\!P \frac{1}{k^2} \mp i\pi\delta(k^2),
\end{equation}
where $I\!\!P$ indicates the principal value. While the imaginary part
is never exposed (for kinematical reasons) in the usual two-to-two
lowest-order partonic scattering amplitudes, in those containing
three-parton correlators it is possible for one internal line to become
soft (along a boundary of the three-body phase space). The three
boundaries of interest are given by the kinematical limits: $x_i\to0$,
where $i=q$, $\bar{q}$ or $g$.

\begin{figure}[htb]
\begin{center}
%==============================================================================%
% T3BLOBS.AXO ((La)TeX input)                                                  %
% Last modified on 25-FEB-1998                     author: Philip G. Ratcliffe %
%==============================================================================%
\begin{picture}(250,50)(0,0)
\multiput(0,0)(150.5,0){2}{
\Line(  0,51)( 20,51) \Line(100,51)( 80,51)
\Line(  0,49)( 20,49) \Line(100,49)( 80,49)
\Line( 11,50)(  8,53) \Line( 11,50)(  8,47)
\Line( 91,50)( 88,53) \Line( 91,50)( 88,47)
\Oval( 50,50)( 10,30)(0)
\Gluon(50,40)( 50,15){2}{4}
}
\Line( 30,42)( 30,15) \Line( 70,42)( 70,15)
\Text( 50,0)[]{(a)}
\Gluon(180,42)(180,15){2}{4} \Gluon(220,42)(220,15){2}{4}
\Text(200,0)[]{(b)}
\end{picture}
\caption{\label{fig:t3blobs}
  The basic three-parton twist-three $qqg$ and $ggg$ hadronic
  amplitudes contributing to transverse-spin asymmetries.}
\end{center}
\end{figure}
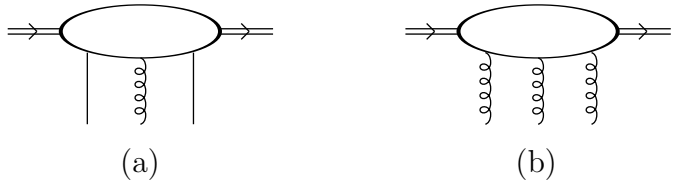

The strong flavour-spin correlation in the measured pion asymmetries
prompts initial consideration of the diagrams of the $qqg$ amplitude
(fig.\,\ref{fig:t3blobs}a). This will certainly demonstrate the full
potential of the approach. However, the triple-gluon correlator
(fig.\,\ref{fig:t3blobs}b) may also contribute \cite{XJ,HES} and should
be taken into account; the technique described here does not depend on
the detailed form of the correlators and thus will suffice in this case
too. Therefore, we shall concentrate on contributions arising from
diagrams of the type shown in fig.\,\ref{fig:t3diags} and, in
particular, on those arising when either a gluon or quark line becomes
soft \cite{ET-1,QS-1}. These may be divided into three classes: gluon
insertion into (\emph{i\/}) initial external lines, (\emph{ii\/}) final
external lines and (\emph{iii\/}) internal lines. We shall consider
these in turn.

The first two classes can, in principle, both provide an imaginary part:
the insertion into an on-shell external line leads to an additional
internal propagator, which may reach the soft limit. However, the
transversity (see later) of the gluon connected to the hadronic
amplitudes in question forces a non-zero transverse momentum in the
struck line. Thus, the collinearity of the initial lines forces such a
contribution to be of even higher twist. On the other hand, the $p_T$
dependence of the final-state parton is just as suggested by the
observed phenomena and only final-state external insertions give
non-vanishing contributions. The last class leads to an imaginary part
only when another external line becomes soft, \ie, when the gluon line
carries all the momentum of the polarised hadron ($x_g=\pm1$). These
diagrams may also be written in a factorised form, viewing them in terms
of soft fermionic insertions; although the final result is more
complicated and both initial- and final-state insertions contribute.

There are two $qqg$ hadronic amplitudes (fig.\,\ref{fig:t3blobs}a) for
the twist-three contribution \cite{PGR-1}:
\begin{equation} \label{eq:blobs}
  D^A(x_1,x_2) \, \gamma_5 \Slashed{p} s_T^\mu
  \qquad \mbox{and} \qquad
  D^V(x_1,x_2) \, \Slashed{p}
  \frac{i\epsilon^{\mu p\bar{p}s_T}}{p.\bar{p}},
\end{equation}
where $p$ and $s_T$ are the momentum and (purely transverse) spin
vectors of the incoming polarised hadron while $\bar{p}$ belongs to the
unpolarised state; typically one takes $p^\mu=E(1,0,0,1)$ and
$\bar{p}^\mu=E(1,0,0,-1)$ in the partonic centre-of-mass frame. The
parton correlators, $D^{A,V}(x_1,x_2)$, have the following symmetry
properties under interchange of their arguments:
\begin{equation} \label{eq:symm}
  D^A(x_1,x_2) \; = \; D^A(x_2,x_1)
  \qquad \mbox{and} \qquad
  D^V(x_1,x_2) \; = \; -D^V(x_2,x_1).
\end{equation}

It is instructive to rewrite the hadron-parton amplitudes using a
suitable helicity basis, in which the calculation simplifies. To do this
we shall adopt a common and convenient notation \cite{MP} and ignore
quark-mass contributions:
\begin{equation}
  u_\pm(p) \; = \; \ket{p\pm}
  \qquad \mbox{and} \qquad
  \bar u_\pm(p) \; = \; \bra{p\pm}.
\end{equation}
We may thus write
\begin{equation}
  \begin{array}{rcl}
   \Slashed{p}
   & = & \ket{p+}\bra{p+} + \ket{p-}\bra{p-},\\
   \gamma_5 \Slashed{p}
   & = & \ket{p+}\bra{p+} - \ket{p-}\bra{p-}.
  \end{array}
\end{equation}
For the amplitudes (\ref{eq:blobs}), the gluon is linearly polarised in
a plane perpendicular to the beam (parallel and orthogonal to
$\vec{s}_T$ respectively for the axial and vector amplitudes). Thus, the
polarisation vectors take the following natural forms:
\begin{equation}
  \xi^\mu_A(p)
  = s_T^\mu
  \qquad \mbox{and} \qquad
  \xi^\mu_V(p)
  = -\frac{i\epsilon^{\mu p\bar{p}s_T}}{p.\bar{p}}.
\end{equation}
A helicity basis may be constructed using these:
\begin{eqnarray}
  \tilde\xi^\mu_\pm(p)
  & = &
  \frac{p.\bar{p} \tilde\eta^\mu + \bar{p}.\tilde\eta p^\mu
   - p.\tilde\eta \bar{p}^\mu \mp i\epsilon^{\mu p\bar{p}\tilde\eta}}
   {2\sqrt{p.\bar{p} \, p.\tilde\eta \, \bar{p}.\tilde\eta}}
  \nonumber
\\
  & = &
  \frac{1}{\sqrt2}
   \left[ s_T^\mu \mp \frac{i\epsilon^{\mu p\bar{p}s_T}}{p.\bar{p}} \right]
  \; = \; \frac{1}{\sqrt2}
   \left[ \xi^\mu_A(p) \pm \xi^\mu_V(p) \strut\right],
\end{eqnarray}
where the choice of auxiliary vector,
\begin{equation}
  \tilde\eta^\mu
  = s_T^\mu + \frac{p^\mu+\bar{p}^\mu}{\sqrt{2p.\bar{p}}}
  \qquad \mbox{with} \qquad
  \tilde\eta^2 = 0,
\end{equation}
implicitly fixes the phase convention for circular polarisation. A more
conventional choice for the phase is to take $\vec\eta$ in the
scattering plane and perpendicular to the beam axis; in terms of such a
set (without the tilde) one has
\begin{eqnarray}
  \tilde\xi^\mu_\pm(p)
  & = &
  e^{\pm i\phi_{s\eta}} \, \xi^\mu_\pm(p),
\end{eqnarray}
where $\phi_{s\eta}$ is the azimuthal angle between $\vec{s}_T$ and
$\vec\eta$.

Expressions (\ref{eq:blobs}) can thus be rewritten as
\begin{equation} \label{eq:blobs1}
  \begin{array}{rl}
   D^A(x_1,x_2) &
   \left[ \ket{p+}\bra{p+} - \ket{p-}\bra{p-} \strut\right]
   \frac{1}{\sqrt2}
   \left[ e^{i\phi} \xi^\mu_+(p) + e^{-i\phi} \xi^\mu_-(p) \right],
\\[6pt]
   D^V(x_1,x_2) &
   \left[ \ket{p+}\bra{p+} + \ket{p-}\bra{p-} \strut\right]
   \frac{1}{\sqrt2}
   \left[ e^{i\phi} \xi^\mu_+(p) - e^{-i\phi} \xi^\mu_-(p) \right].
  \end{array}
\end{equation}
Note that, since $\xi_-=\xi_+^*$, the last factors in the two
expressions above are respectively purely real and purely imaginary. One
also clearly sees how the axial (vector) contributions are related to
amplitudes involving quark (gluon) helicity differences. The necessary
phases are generated by combinations of the propagator imaginary parts
and the gluon polarisation-vector phases.

The triple-gluon amplitudes have been considered by Ji \cite{XJ} and
lead to more complex expressions involving a number of correlation
functions. However, the common simplifying characteristic is that the
associated gluon polarisation projectors are restricted to the
transverse plane and so can be represented by physical polarisation
vectors.

%------------------------------------------------------------------------------%
\section{Factorisation in single-spin \boldmath{$\tau\!=\!3$} amplitudes}
\label{sec:factor}

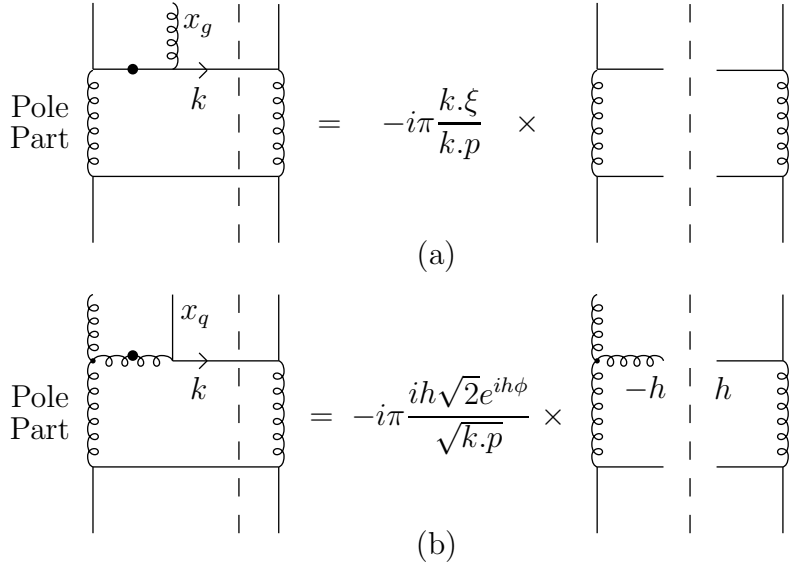
\begin{figure}[htb]
\begin{center}
%==============================================================================%
% QFACTOR.AXO ((La)TeX input)                                                  %
% Last modified on 23-JAN-1998                     author: Philip G. Ratcliffe %
%==============================================================================%
\begin{picture}(280,210)(0,0)
\multiput(0,0)(0,110){2}{
\Text  (  0, 62)[]{Pole}
\Text  (  0, 50)[]{Part}
\Line  ( 90,100)( 90, 75)
\Gluon ( 20, 75)( 20, 35){-2}{6} \Gluon( 90, 75)( 90, 35){2}{6}
\Line  ( 20, 35)( 20, 10)        \Line ( 90, 35)( 90, 10)
\Line  ( 20, 35)( 90, 35)
\DashLine( 75,100)( 75,10){7}
\Line  ( 60, 72)( 63, 75)        \Line ( 60, 78)( 63, 75)
\Text  ( 60, 65)[]{$k$}
\Line  (280,100)(280, 75)
\Gluon (210, 75)(210, 35){-2}{6} \Gluon(280, 75)(280, 35){2}{6}
\Line  (210, 35)(210, 10)        \Line (280, 35)(280, 10)
\Line  (255, 75)(280, 75)
\Line  (210, 35)(235, 35)        \Line (255, 35)(280, 35)
\DashLine(245,100)(245,10){7}
}
\Line  ( 20,210)( 20,185)        \Gluon( 50,210)( 50,185){2}{4}
\Line  ( 20,185)( 90,185)
\Vertex( 35,185){2}
\Text  (105,165)[l]{$\displaystyle=
                     \quad-i\pi\frac{k.\xi}{k.p}
                     \quad\times$}
\Text  ( 54,202)[]{$\quad x_g$}
\Line  (210,210)(210,185)
\Line  (210,185)(235,185)
\Text  (150,115)[]{(a)}
\Gluon ( 20,100)( 20, 75){-2}{4}  \Line ( 50,100)( 50, 75)
\Gluon ( 20, 75)( 50, 75){2}{4}   \Line ( 50, 75)( 90, 75)
\Vertex( 20, 75){1}
\Vertex( 35, 77){2}
\Text  (102, 55)[l]{$\displaystyle=
                     \;-i\pi\frac{ih\sqrt2e^{ih\phi}}{\sqrt{k.p}}
                     \;\times$}
\Text  ( 53, 92)[]{$\quad x_q$}
\Gluon (210,100)(210, 75){-2}{4}
\Gluon (210, 75)(235, 75){2}{4}
\Vertex(210, 75){1}
\Text  (237, 65)[r]{$-h$}
\Text  (255, 65)[l]{$h$}
\Text  (150,  5)[]{(b)}
\end{picture}
\caption{\label{fig:qfactor}
  Graphical representation of the amplitude factorisation in the case
  of soft external (a) gluon and (b) quark lines. The solid circle
  indicates the line from which the imaginary piece is extracted, and
  $\xi$ refers to the gluon entering the factorised vertex.}
\end{center}
\end{figure}

Let us consider first of all the case of soft-gluon insertions into
external quark lines, as in the left-hand diagram of
fig.\,\ref{fig:t3blobs}a. Extracting the imaginary part of the quark
line (marked $\bullet$ in the figure) to the left of the gluon vertex
forces $x_g=0$; taking this into account, the vertex may be written as
\begin{equation} \label{eq:gfactor}
  \xi^\mu_X(p) \bra{k,h_k} \gamma_\mu \slashed{k} \dots
  = \bra{k,h_k} \slashed{\xi}_X \sum_h \ket{k,h} \bra{k,h} \dots
  \qquad (X = A,V),
\end{equation}
where the ellipsis indicates the rest of the amplitude to the left of
the vertex, and colour factors have been suppressed. Including the
remnant factors from the imaginary propagator part and factoring the
$\bra{k,h}$ projector above into the rest of the amplitude,
eq.\,(\ref{eq:gfactor}) reduces to a simple factor:
\begin{equation}
  -i\pi \frac{k.\xi_X(p)}{k.p} \delta(x_g),
\end{equation}
multiplying the now pure two-to-two amplitudes (see the right-hand
diagram of fig.\,\ref{fig:qfactor}a).
The complex-conjugate diagrams acquires a minus sign, arising from the
opposite sign of the $i\varepsilon$ in the propagator.

Soft-gluon insertions into external gluon lines lead to expressions of
the type:
\begin{equation}
  \sum_{\lambda}
  V_{\mu\sigma\nu}
  \xi_X^\mu(p)
  \xi_{\lambda}^{*\sigma}(k)
  \xi_{\lambda_k}^\nu(k)
  \xi_{\lambda}^\rho(k) \dots,
\end{equation}
where the rightmost \emph{circular\/} gluon polarisation vector will be
factored into the remaining amplitude (represented by the ellipsis),
and $V_{\mu\sigma\nu}$ is just the three-gluon vertex here:
\begin{equation}
  V_{\mu\sigma\nu}
  = g_{\mu\sigma}(p-k)_\nu
  + g_{\nu\mu}(-k-p)_\sigma
  + g_{\sigma\nu}2k_\mu.
\end{equation}
Only the last term survives (owing to the gauge choice) and we obtain
\begin{equation}
  -i\pi \frac{k.\xi_X(p)}{k.p} \delta(x_g) \delta_{\lambda,-\lambda_k},
\end{equation}
which has the same structure as the previous case, except that the
gluon helicity is flipped ($\lambda=-\lambda_k$). And with the phase
conventions adopted one has
\begin{equation}
  k.\xi_\pm(p)
  = {\textstyle\frac{1}{\sqrt2}} |k_T| e^{\pm i\phi_{k\eta}},
\end{equation}
where $\phi_{k\eta}$ is the azimuthal angle between $\vec{k}_T$ and
$\vec\eta$. The particular phase dependence on $\phi_{k\eta}$ is just
what is needed: in combination with that coming from the initial state
gluon ($\phi_{s\eta}$, see above), it leads to the expected
$\sin\phi_{ks}$ dependence of the final cross-section.

Three selection rules emerge:
\begin{enumerate}
\item The transverse nature of the gluon kills all contributions of
  initial-state insertions ($k=p$ or $\bar{p}$). Note that, for
  insertions into the incoming lines from the other (unpolarised)
  hadron, this depends on the choice of $p$ as the gauge-fixing vector
  for the gluons from the other hadron.
\item Unless the second hadron is also polarised, the $qqg$ axial
  contribution vanishes owing to parity conservation, as it is
  proportional to a helicity difference for the incoming quark from
  the first hadron.
\item Although proportional to a quark helicity sum, the $qqg$ vector
  contribution does not survive as it is multiplied by $D^V(x,x)$,
  which vanishes according to eq.\,\ref{eq:symm}.\footnote{We shall
  comment later on the possible contribution of higher-order poles.}
\end{enumerate}
Note also that the axial contribution, were it non-vanishing, would lead
to a $\cos\phi$ dependence, \ie, to an up-down asymmetry.

It is possible to treat the case of soft external quark lines similarly,
as in the left-hand diagram of fig.\,\ref{fig:qfactor}b. For want of
better terminology, we shall call these \emph{soft-quark\/} insertions;
although a description in terms of insertion would be more pertinent to
the case of a supersymmetric theory. The only subtlety is the change in
nature of the resulting external particle: a fermionic insertion changes
a fermion to a boson and \emph{vice versa}. The imaginary piece of the
gluon line to the left of the vertex forces $x_q=0$; taking this into
account and explicitly including the effective soft-quark spinor, the
vertex may be written as
\begin{equation} \label{eq:qfactor}
  \sum_\lambda \bra{k,h_k} \gamma_\mu \ket{p,h} \; \xi^\mu_\lambda(k) \;
  \xi^{\nu*}_\lambda(k) \dots,
\end{equation}
where again the rightmost term will be factored into the remaining
amplitude. Including the various factors from the denominator \etc,
eq.\,(\ref{eq:qfactor}) reduces to:
\begin{equation} \label{eq:qfactor1}
  -\frac{i \pi}{k.p}\delta(x_q) \cdot
  i h \sqrt{2k.p} \; e^{i h \phi} \delta_{\lambda,-h},
\end{equation}
where the factored gluon polarisation vector carries helicity $-h$
(see the right-hand diagram of fig.\,\ref{fig:qfactor}b). Here the
selection rule excluding initial-state insertions applies only to the
partons from the same hadron.

We also see that both the axial and vector structures may contribute
here, as they are proportional to $D^{A,V}(0,x)$. Moreover, the
well-known helicity-conservation rules (forbidding the so-called
maximally violating amplitudes \cite{MP,BDK}) force the non-zero
contributions to come only from the terms in eq.\,(\ref{eq:blobs1}) with
$(h_q,\lambda_g)=(\pm,\mp)$. Thus, the axial and vector contributions
arise in simple linear combinations:
\begin{equation} \label{eq:dplusminus}
  D^A(0,x) \pm D^V(0,x) = D^\pm(0,x) = \mp D^\mp(x,0),
\end{equation}
see ref.\,\cite{PGR-1} for the relevant definitions. There only remains
to calculate the case of insertions where the gluon is the external line
and the quark, internal. This is, however, simply the complex conjugate
of factor (\ref{eq:qfactor1}).

It is worth making a few further observations. Factorisation of the
amplitudes immediately clarifies the possibility of large asymmetries,
where once they were believed to be suppressed. First of all, the colour
overlap is only slightly modified while the phase-space is unaltered,
and thus little is lost for reasons of mismatch; the (supersymmetric
\cite{MP,BDK}) Ward identities guarantee the close similarity between
amplitudes where a fermion line is replaced by a gluon. Indeed, the
interference is not between differing kinematical configurations (as
often found in early analyses) but simply between spin-flip and non-flip
amplitudes; the quark-insertion factor shown in eq.\,(\ref{eq:qfactor1})
explicitly displays the spin-flip nature (between quark and gluon).

In the above we have ignored the possibility, discussed in the
literature \cite{Pion}, that the correlator $D^V(x_1,x_2)$ might be
accompanied by an extra pole in $(x_1-x_2)$.\footnote{The author is
particularly indebted to Oleg Teryaev for clarifying discussions on this
point.} Should this prove to be the case, then the requirement of an
imaginary part would still force the $\delta$-function from the
propagator. A Taylor expansion of $D^V(x_1,x_2)$ about the point
$(x_1-x_2)=0$ would pick out the first derivative of the correlator but
leave all other algebraic manipulations as before. Thus, the selection
rule excluding terms in $D^V$ would be avoided while the factorisation
property would remain unaltered.

Finally, the apparent higher order in $\alpha_s$ of the diagrams is
removed by the absorption of the gluon propagator and vertices into the
hadronic blob itself (as dictated by gauge invariance), leaving an
\emph{effective\/} tree-level leading-order graph. Moreover, the
expressions may now be written in compact form and require little effort
to calculate; all two-to-two pQCD amplitudes are already well known.
Only the slightly modified colour factors remain to be evaluated, a task
easily performed with the aid of a symbolic manipulation programme.

%-------------------------------- Conclusions ---------------------------------%
\section{Conclusions}
\label{sec:conclusions}

The resulting forms of the amplitudes given above greatly simplify the
calculation of the asymmetries: the calculation of the tens of Feynman
diagrams normally contributing is reduced to the evaluation of products
of the simple factors derived here and known two-body helicity
amplitudes. Since all two-body helicity amplitudes have indeed already
been calculated in pQCD we shall merely present formal expressions for
the asymmetries, as sums over a very limited number of amplitudes for
fixed helicities. The soft-insertion factorisation thus allows the
partonic cross-section to be expressed in the following compact form:
\begin{eqnarray}
  \Delta\hat\sigma
  & = &
  \sum_{i,j} C_{ij}
   \, {\cal M}_i        (x,\bar{x},k_T)
   \, {\cal M}_j^\dagger(x,\bar{x},k_T) ,
\end{eqnarray}
where $C_{ij}$ represents both the insertion factors given above and
modified colour factors, and the ${\cal{M}}_i$, the individual two-body
amplitudes. This much simplified form is ideal for the development of a
computer programme (\eg, MadGraph \cite{SL}) based on helicity-amplitude
subroutines (\eg, Helas \cite{MWH}) for the automatic generation of
cross-sections for any twist-three single-spin asymmetry.

In concluding, let us first of all highlight a difference in the
interpretation of the origin of the $x_F$ dependence with respect to
ref.\,\cite{QS-1}, where the presence of the derivative of a $qqg$
correlator was claimed responsible for the rise in polarisation effects
towards the edges of parton phase-space. Here, in contrast, the remnant
factors of $(-t)^{-\frac12}$, $(-u)^{-\frac12}$ are seen to lie at the
origin of this behaviour. It should be stressed that this transparency
is due to the factorisation procedure presented.

It is also worth pointing out that the triple-gluon contributions, being
insensitive to flavour, are also suggested by the experimentally
observed approximately equal magnitudes and opposite signs of the
$\pi^+$ and $\pi^-$ asymmetries, where one might have expected a ratio
of the order of three to one (with opposite signs), according to SU(6).
The (flavour-blind) triple-gluon contribution could lead to just the
required net shift of both asymmetries in the same direction.

With the above formulation in terms of four-body amplitudes, it will not
be difficult to set up an analysis of the existing data, from which a
general parametrisation of the partonic correlators may be determined in
a manner similar to that of Anselmino \etal.\ \cite{ABM}. On the other
hand, the procedure adopted here is purely pQCD based and, in
particular, requires no assumptions as to the nature of intrinsic
$s_T$-$k_T$ correlations. Indeed, the factorisation property presented
should help in clarifying the physical significance of the trade-off
between the operator-product expansion description in terms of fields
with only ``good'' components \cite{RLJ} and the $k_T$ dependence
augmenting the parton picture \cite{ABM}.

As an example process, we have considered left-right asymmetries for
final-state hadrons produced in hadron-hadron collisions with a single
initial state polarised. However, it is clear that the proposed
factorisation may be extended to many other processes in
straight-forward manner, including those involving polarised and
unpolarised twist-three fragmentation functions. As remarked above, one
could also consider measuring the up-down asymmetry predicted to exist
for scattering involving one transverse polarisation and one
longitudinal. While this asymmetry also contains twist-2 contributions,
it would allow for a cross-check measurement of some of the
distributions invoked here. The obvious advantage of the single-spin
measurements (apart from their experimental accessibility) lies in their
automatic and complete filtering of all twist-2 effects.

%------------------------------ Acknowledgments -------------------------------%
\section{Acknowledgments}

The author wishes to express thanks to Bob Jaffe and George Sterman, for
rekindling his interest in this subject and for useful discussions; and
to Oleg Teryaev for many illuminating discussions, his encouraging
comments and critical reading of the manuscript. Gratitude is also due
to the Riken-BNL centre for hospitality during the germinal period of
this study.

%\bibliography{pigrobib}

\begin{thebibliography}{99}

\bibitem{DataG2} D.L.~Adams \emph{et~al.\/} (FNAL-E704 collab.),
  \emph{Phys.\ Lett.\/} \textbf{B264} (1991)~462; \\
  D.L.~Adams \emph{et~al.\/} (FNAL-E581 collab.), \emph{Z. Phys.\/}
  \textbf{C56} (1992)~181; \\
  A.~Bravar \emph{et~al.\/} (FNAL-E704 collab.),
  \emph{Phys.\ Rev.\ Lett.\/} \textbf{75} (1995)~3073.

\bibitem{KPR} G.L.~Kane, J.~Pumplin and W.~Repko,
  \emph{Phys.\ Rev.\ Lett.\/} \textbf{41} (1978)~1689.

\bibitem{Twist3} E.V.~Shuryak and A.I.~Vainshtein,
  \emph{Nucl.\ Phys.\/} \textbf{B199} (1982)~451; \\
  A.P.~Bukhvostov, \'E.A.~Kuraev and L.N.~Lipatov,
  \emph{Sov.\ Phys.\ JETP\/} \textbf{60} (1984)~22; \\
  A.P.~Bukhvostov, \'E.A.~Kuraev, L.N.~Lipatov and G.V.~Frolov,
  \emph{Nucl.\ Phys.\/} \textbf{B258} (1985)~601; \\
  A.V.~Efremov and O.V.~Teryaev, \emph{Sov.\ J. Nucl.\ Phys.\/}
  \textbf{39} (1984)~962.

\bibitem{ET-1} A.V.~Efremov and O.V.~Teryaev, \emph{Yad.\ Fiz.\/}
  \textbf{39} (1984)~1517; \emph{Phys.\ Lett.\/} \textbf{B150}
  (1985)~383.

\bibitem{PGR-1} P.G.~Ratcliffe, \emph{Nucl.\ Phys.\/} \textbf{B264}
  (1986)~493.

\bibitem{RLJ} R.L.~Jaffe, \emph{Comm.\ Nucl.\ Part.\ Phys.\/}
  \textbf{14} (1990)~239.

\bibitem{DataSSA} K.~Abe \emph{et~al.\/} (E143 collab.),
  \emph{Phys.\ Rev.\ Lett.\/} \textbf{76} (1996)~587; \\
  K.~Abe \emph{et~al.\/} (E154 collab.), \emph{Phys.\ Lett.\/}
  \textbf{B404} (1997)~377; \\
  D.~Adams \emph{et~al.\/} (SM collab.), \emph{Phys.\ Lett.\/}
  \textbf{B336} (1994)~125.

\bibitem{KH} K.~Heller, in the proc.\ of
  \emph{The XII Int.\ Symp.\ on High Energy Spin Physics\/}
  (Amsterdam, Sept.\ 1996), eds.\ C.W.~de~Jager, T.J.~Ketel,
  P.J.~Mulders, J.E.J.~Oberski and M.~Oskam-Tamboezer
  (World Sci., 1997), p.~23.

\bibitem{ABM} M.~Anselmino, M.~Boglione and F.~Murgia,
  \emph{Phys.\ Lett.\/} \textbf{B362} (1995)~164.

\bibitem{OVT} O.V.~Teryaev, JINR preprint, e-print hep-ph/9808335
  (1998).

\bibitem{PGR-2} P.G.~Ratcliffe, in the proc.\ of
  \emph{The 6th.\ Int.\ Symp.\ on High Energy Spin Physics\/}
  (Marseille, Sept.\ 1984), ed.\ J.~Soffer; \emph{J. de Phys.\ Suppl.\/}
  \textbf{46} (C2) (1985)~31.

\bibitem{QS-1} J.~Qiu and G.~Sterman, \emph{Phys.\ Rev.\ Lett.\/}
  \textbf{67} (1991)~2264; \emph{Nucl.\ Phys.\/} \textbf{B378} (1992)~52.

\bibitem{KT} V.M.~Korotkiian and O.V.~Teryaev, \emph{Phys.\ Rev.\/}
  \textbf{D52} (1995)~4775.

\bibitem{HES} N.~Hammon, B.~Ehrnsperger and A.~Sch{\"a}fer,
  \emph{J. Phys.\/} \textbf{G24} (1998)~991.

\bibitem{Pion} A.V.~Efremov, V.M.~Korotkiyan and O.V.~Teryaev,
  \emph{Phys.\ Lett.\/} \textbf{B348} (1995)~577;\\
  J.~Qiu and G.~Sterman, SUNY preprint ITP-SB-98-28, e-print
  hep-ph/9806356 (1998).

\bibitem{Drell-Yan} N.~Hammon, O.~Teryaev and A.~Sch{\"a}fer,
  \emph{Phys.\ Lett.\/} \textbf{B390} (1997)~409;\\
  D.~Boer, P.J.~Mulders and O.V.~Teryaev, \emph{Phys.\ Rev.\/}
  \textbf{D57} (1997)~3057.

\bibitem{XJ} X.~Ji, \emph{Phys.\ Lett.\/} \textbf{B289} (1992)~137.

\bibitem{MP} M.L.~Mangano and S.J.~Parke, \emph{Phys.\ Rep.\/}
  \textbf{200} (1991)~301.

\bibitem{BDK} Z.~Bern, L.~Dixon and D.A.~Kosower,
  \emph{Ann.\ Rev.\ Nucl.\ Part.\ Sci.\/} \textbf{46} (1996)~109.

\bibitem{SL} T.~Stelzer and W.F.~Long, \emph{Comp.\ Phys.\ Comm.\/}
  \textbf{81} (1994)~357.

\bibitem{MWH} H.~Murayama, I.~Watanabe and K.~Hagiwara, KEK preprint
  KEK-91-11 (1991).

\end{thebibliography}
%\bibliographystyle{pigrobib}
%\end{document}

%-------------------------------- Bibliography --------------------------------%

%----------------------------------- The End ----------------------------------%
\end{document}